\documentclass[sigconf]{acmart}

\usepackage{enumitem}

\newcommand{\RQ}[2]{\begin{description}[leftmargin=0cm]
  \item[#1] \textit{#2}
  \end{description}}
\newcommand{\IP}{\textbf{(ip)} }
\newcommand{\IR}{\textbf{(ir)} }

\setcopyright{none}
\copyrightyear{2024}
\acmYear{2024}
\acmDOI{}

\acmConference[ESEM '24 Registered Reports]{18th ACM/IEEE International Symposium on Empirical Software Engineering and Measurement}{October 20--25, 2024}{Barcelona, Spain}
\acmISBN{}

\begin{document}

\include{response}

\title[Local Software Buildability across Java Versions]{Local Software Buildability across Java Versions\\ (Registered Report)}

\author{Mat\'u\v{s} Sul\'ir}
\email{matus.sulir@tuke.sk}
\orcid{0000-0003-2221-9225}
\affiliation{%
  \institution{Technical University of Ko\v{s}ice}
  \city{Ko\v{s}ice}
  \country{Slovakia}
}

\author{Jaroslav Porub\"an}
\email{jaroslav.poruban@tuke.sk}
\orcid{0000-0001-9706-2897}
\affiliation{%
  \institution{Technical University of Ko\v{s}ice}
  \city{Ko\v{s}ice}
  \country{Slovakia}
}

\author{Sergej Chodarev}
\email{sergej.chodarev@tuke.sk}
\orcid{0000-0002-9293-0859}
\affiliation{%
  \institution{Technical University of Ko\v{s}ice}
  \city{Ko\v{s}ice}
  \country{Slovakia}
}

\renewcommand{\shortauthors}{Sul\'ir et al.}

\begin{abstract}
\textbf{Context:} Downloading the source code of open-source Java projects and building them on a local computer using Maven, Gradle, or Ant is a common activity performed by researchers and practitioners. Multiple studies so far found that about 40--60\% of such attempts fail. Our experience from the last years suggests that the proportion of failed builds rises continually even further. \textbf{Objective:} First, we would like to empirically confirm our hypothesis that with increasing Java versions, the percentage of build-failing projects tends to grow. Next, nine supplementary research questions are proposed, related mainly to the proportions of failing projects, universal version compatibility, failures under specific JDK versions, success rates of build tools, wrappers, and failure reasons. \textbf{Method:} We will sample 2,500 random pure-Java projects having a build configuration file and fulfilling basic quality criteria from GitHub. We will try to automatically build every project in containers with Java versions 6 to 23 installed. Success or failure will be determined by exit codes, and standard output and error streams will be saved. A majority of the analysis will be performed automatically using reproducible scripts. 
\end{abstract}

% The code below is generated by the tool at http://dl.acm.org/ccs.cfm.
\begin{CCSXML}
<ccs2012>
   <concept>
       <concept_id>10011007.10011006.10011041</concept_id>
       <concept_desc>Software and its engineering~Compilers</concept_desc>
       <concept_significance>500</concept_significance>
       </concept>
   <concept>
       <concept_id>10011007.10011006.10011071</concept_id>
       <concept_desc>Software and its engineering~Software configuration management and version control systems</concept_desc>
       <concept_significance>300</concept_significance>
       </concept>
   <concept>
       <concept_id>10011007.10011006.10011008.10011009.10011011</concept_id>
       <concept_desc>Software and its engineering~Object oriented languages</concept_desc>
       <concept_significance>300</concept_significance>
       </concept>
 </ccs2012>
\end{CCSXML}

\ccsdesc[500]{Software and its engineering~Compilers}
\ccsdesc[300]{Software and its engineering~Software configuration management and version control systems}
\ccsdesc[300]{Software and its engineering~Object oriented languages}

\keywords{build tools, Java, Gradle, Maven, Ant, compatibility}

\maketitle

%%%%%%%%%%%%%%%%%%%%%%%%%%%%%%%%%%%%%%%%%%%%%%%%%%%%%%%%%%%%

\section{Introduction}
\label{s:introduction}

Building an open-source software project locally from the source code is a fundamental precondition to many frequent activities performed by researchers, students, and practitioners. Software engineering researchers often utilize executable programs paired with their original source code as study objects --- e.g., for controlled experiments with human participants or for studies depending on bytecode or dynamic analysis. Industrial programmers sometimes need to build a third-party program or library, mainly to find the root cause of a bug, inspect a system thoroughly, or modify it to suit the business requirements. Students may be asked to download and build a software system as a part of a course assignment.

The model situation is as follows: A person downloads a project via Git (or as a source code archive file) locally. On the given computer, usually only one JDK (Java Development Kit) version is installed, commonly the latest LTS (long-term support) version or the latest version at all. Running the supplied Maven or similar build script, either in a console or an IDE (integrated development environment), results in a failure with a lengthy and incomprehensible error message. After a few unsuccessful attempts to fix it, the person gives up. If it makes sense in the given context, the person downloads an alternative project, only to find out the compilation results in an error too.

\subsection{Related Work}
\label{s:related}

The phenomenon of build failures in general has been the subject of many empirical studies. In a nutshell, the unique property of our work is that we would like to study local (not continuous integration) build failures on a large sample (thousands, not tens) of open-source projects using multiple versions (not a single version) of JDK.

\subsubsection{Continuous Integration Builds}
Several studies are focused on CI builds. Rausch et al. \cite{Rausch17empirical} analyzed Travis CI build logs of 14 open-source projects. Their fail ratio ranged from 14\% to 69\%, and the builds most frequently failed because of tests. From the point of view of the projects' history, most build errors occurred in consecutive snapshots. Kerzazi et al. \cite{Kerzazi14why} analyzed build errors in one specific software company and measured a 17.9\% build breakage ratio. Based on the interview results, many developers are unaware of server build breakages and fix them multiple hours after they occur. Seo et al. \cite{Seo14programmers} found that about 37\% of C++ and 29.7\% of Java cloud-based build processes at Google failed. Zolfagharinia et al. \cite{Zolfagharinia19study} studied CI build failures of Perl software systems. They included a comparison of failures produced by different Perl versions and operating systems (OSs). Most notably, 17\% of project--OS pairs succeed on some Perl versions but fail on other ones. The most common failure reason was a dependency error. Another study focused specifically on the long duration of CI builds \cite{Ghaleb19empirical}; 40\% of the analyzed builds were longer than 30 minutes. In a study focused specifically on compiler errors \cite{Zhang19large}, the most common error type was unsuccessful symbol resolution.

CI builds, however, differ significantly from local builds since they run in a controlled environment precisely specified by configuration files. In addition, not all projects include CI configuration.

\subsubsection{Local Builds}
\label{s:local}
Among the studies focused on local builds, some were limited in scope or scale. For example, Neitsch et al.~\cite{Neitsch12build} qualitatively explored build system issues in five selected multi-language software projects. They described multiple build anti-patterns, such as Filename Collision or Ignored Error. Rabbani et al.~\cite{Rabbani18revisiting} analyzed developers' intermediate builds, mainly of their own projects, in the Visual Studio IDE. About 13\% of them ended by failure. Horton and Parnin~\cite{Horton18gistable} evaluated the executability of short code snippets called Gists. Since they usually do not include a list of dependencies in a standardized form, less than a quarter of them were executable by default.

A few large-scale studies focused on local builds of open-source projects. Tufano et al. \cite{Tufano17there} tried to build past snapshots (commits) of 100 Java projects. They found that on average, 62\% of commits per project could not be compiled. In a replication by Maes-Bermejo et al. \cite{Maes-Bermejo22revisiting}, which overcame the limitation only to Maven, a mean of 65\% of commits per project were uncompilable.

In 2016, Sul\'ir and Porub\"an \cite{Sulir16quantitative} tried to build the latest commits of more than 7,200 open-source Java projects from GitHub using JDK 8. More than 38\% of projects failed to be built. Maven configurations succeeded the most often and Ant scripts had the highest failure ratio. The study was replicated in 2020 \cite{Sulir20large} on a new set of projects, using JDK 11 and newer versions of build tools installed. The proportion of failures increased to 59\%. In both executions, dependency resolution and Java compilation errors were among the most frequent failure reasons.

None of the studies, however, tried to compile a large number of projects using multiple JDK versions.

\subsubsection{Java API Incompatibility}
One of the reasons for build failures is API incompatibility. There are studies that specifically focus on this topic. For instance, Jezek et al. \cite{Jezek15how} studied the breakage of Java APIs in general. They found that although incompatible API changes in libraries are very common, they rarely cause actual problems in the depending programs. Dietrich et al. \cite{Dietrich16what} surveyed the developers' knowledge about binary, source, and behavioral compatibility of the Java language and virtual machines. According to their study, many developers lack the knowledge of differences between these types of compatibility and the rules that apply. Sawant et al. \cite{Sawant18reaction} analyzed API deprecations in Maven Central. They found, for example, that few clients update the API versions they use, and incompatible changes are often solved by a deletion of the affected call instead of its suitable replacement. In our work, though, we suppose the build failures will occur also because of reasons other than API incompatibility.

\subsection{Aim}
\label{s:aim}

We will try to fully automatically build 2,500 open-source Java projects from GitHub, using their Maven, Gradle, or Ant script. Every project will be built 18 times, each time in a container with a different JDK version installed, ranging from 6 to 23.

The number of projects was determined as a compromise between statistical confidence and resource consumption/feasibility.

We aim to state that a particular proportion of the projects fail under a certain JDK with a 95\% confidence level, a 0.5 estimated build failure proportion (the most conservative choice) and a 2\% absolute margin of error. According to the simple random sample size calculation by Levy and Lemeshow \cite{Levy08sampling}, the minimum sample is 2,191 projects for an estimated population of 25,000 projects or 2,401 for an unknown population size. We decided to round the latter number up to 2,500 projects.

Based on our experience from the previous studies \cite{Sulir16quantitative,Sulir20large}, on a virtual machine with 8 CPU cores and 16 GB RAM, we can build about 1,000 projects per day. If we build each project 18 times, we can expect the execution to end in approximately 6.5 weeks sequentially, or less with parallelization.

The Java ecosystem is particularly interesting in this aspect for two reasons. First, it is one of the most popular languages, continually appearing in the first places in ratings. Second, Java was previously famous for its excellent stability and compatibility across versions. Java 9 brought a number of breaking changes,\footnote{\url{https://www.oracle.com/java/technologies/javase/9-relnotes.html}} most notably modularization (Project Jigsaw). Furthermore, the release cadence significantly increased from that moment, and a new major version is released every six months. Practically each of them removes multiple deprecated APIs (application programming interfaces). In Java 17, illegal reflective access was definitively blocked,\footnote{\url{https://www.oracle.com/java/technologies/javase/17-relnote-issues.html}} which poses another significant compatibility problem.

The rationale behind the specific version range is that JDK versions 5 and lower were distributed under a proprietary license prohibiting certain kinds of redistribution,\footnote{\url{https://www.oracle.com/java/technologies/java-archive-javase5-downloads.html}} which could cause problems when creating a replication package for our study. JDK 23 will be the latest generally available Java version at the moment of the actual study execution.

\subsection{General Implications}
\label{s:implications}

Both researchers and practitioners are often frustrated by failing local builds of open-source projects. Thoroughly examining the relationship between failing builds and JDK versions would, among other things:
\begin{itemize}
\item quantify the severity of the JDK incompatibility problem,
\item inform the developers about what failure rates they should expect in particular situations,
\item suggest to project owners techniques they could use to achieve universal compatibility across Java versions,
\item provide a reminder for language designers about incompatible language changes that are ill-received by the community,
\item provide hints to developers regarding the success rates of the individual build tools and the potential advantage of using wrapper scripts,
\item and suggest which build fail reasons should the researchers prioritize when designing automated build repair approaches.
\end{itemize}

Finally, the dataset we would produce is also a contribution in itself. It would offer a unique and rich source of information about the build process across many JDK versions, allowing the researchers to answer other interesting questions not proposed in this report.

%%%%%%%%%%%%%%%%%%%%%%%%%%%%%%%%%%%%%%%%%%%%%%%%%%%%%%%%%%%%

\section{Research Questions and Motivation}
\label{s:rqs}

In this section, we will state the hypothesis of our predominantly confirmatory study and research questions that supplement it (with only RQ4 being purely exploratory). For each of them, implications for practitioners \textbf{(ip)} and implications for researchers \textbf{(ir)} acting as motivation will be mentioned.

Our main hypothesis is:

\RQ{H}{With increasing JDK version numbers, the proportion of build-failing projects tends to grow.}

\IP Java was known as an archetype of compatibility among developers. The official documentation for former JDK versions tended to say they are ``strongly compatible with previous versions of the Java platform'' and ``almost all existing programs should run \dots without modification.''\footnote{For example, \url{https://www.oracle.com/java/technologies/javase/8-compatibility-guide.html}.} We would like to empirically confirm the intuition of developers that with newer Java versions, this is no longer true. \IR The confirmation of this hypothesis would show that Java version incompatibility is a serious research problem that deserves attention. It also acts as a motivation to answer the research questions proposed in this section.

We propose the following general question:

\RQ{RQ1}{What proportion of projects is buildable by their supplied script, using Java Development Kit versions ranging from 6 to 23?}

\IP \IR This is an umbrella question that more precisely quantifies the result of our hypothesis. We will provide implications for each detailed question separately.

The build failure problem can be perceived from three main aspects --- projects, JDKs, and build tools.

\subsection{Projects} \label{s:projects}

First, the project-related questions are:

\RQ{RQ2}{What proportion of projects failing under the latest available JDK can be built using an earlier JDK?}

\IP Here we suppose a developer has the latest Java version available at the given time (JDK 23) installed, and we would like to find out how many of the failing builds can be fixed just by choosing another JDK. This quantifies to what degree is the strategy of trying older JDKs worth trying in practice. \IR The determined percentage empirically defines the maximum proportion of projects for which automated build repair tools working by switching from the latest JDK to one of the older ones could be useful.

\RQ{RQ3}{What proportion of projects always fails, passes only for some JDKs, and passes for all JDKs?}

\IP The answer to this question will suggest to the developers how (un)reasonable is to download an unknown project from GitHub and expect it to build successfully --- either using one JDK or even if all JDKs are tried. \IR This will allow the researchers to obtain more general and up-to-date information about the prevalence of build failures in open-source projects than the existing studies mentioned in section~\ref{s:local}. Knowing that a certain percentage of projects does not build regardless of the JDK version helps focus research in the area of automated build repair also on methods unrelated to version compatibility.

\RQ{RQ4}{How do projects achieve compatibility with all JDKs at once?}

\IP After filtering out fairly trivial projects, the qualitative answer to this question may act as an inspiration for project maintainers on how to achieve compatibility across all JDK versions. We suppose, for example, some of them download the necessary versions of language toolkits from the Internet (making the user's installed JDK version irrelevant). However, we might also find other techniques, such as using a restricted subset of the Java language. \IR The found set of projects could be used by programming language researchers for specific scenarios that require building or running the same code across a wide range of Java versions, e.g., for benchmarking.

\subsection{JDKs}

Next, the questions concerning the compatibility of individual JDK versions are:

\RQ{RQ5}{Which JDK version can build the largest/smallest number of projects?}

\IP When combining many libraries in one large software project, there are sometimes conflicts between Java version ranges supported by each library. If the percentage of projects buildable using at least one JDK (from \textbf{RQ3}) considerably differs from the percentage of projects buildable under the single most ``powerful'' JDK, it means the ecosystem is fragmented across incompatible versions. Thanks to this RQ, practitioners could know which JDK version to use when their goal is to aim for the largest compatibility. \IR In software repository mining research, it is sometimes necessary to have a dataset of buildable software projects. Traditionally, the whole set should be compilable under any recent Java version. Due to version fragmentation, this might be no longer straightforward, so we would like to suggest the best Java version (or a set of versions) to use.

\RQ{RQ6}{Which JDK versions cause sudden drops in the passing rate?}

\IP \IR Not all changes in programming languages are well-received by the community. Sudden drops in passing rates should signify versions introducing severe incompatibilities largely ignored by the developers. Although we already described such examples in section~\ref{s:aim} (modularization, blocking illegal reflective access), the actual results might be surprising. We will summarize the list of main incompatible changes in the given Java versions, which will act as a reminder for language designers.

\subsection{Build Tools} \label{s:tools}

Finally, sub-questions related to the build tools are:

\RQ{RQ7}{What proportion of projects can be built using individual JDKs, per each build tool (Maven, Gradle, Ant)?}

\IP This could tell the developers what should be their expectations when trying to build a project using any of the mentioned build tools. \IR For researchers, this quantifies the necessity of designing automated build repair approaches for individual build tools.

\RQ{RQ8}{Are there differences in failure rates between the projects utilizing wrappers (e.g., the \texttt{gradlew} script located directly in the repository) and the projects requiring system-wide build tool installation?}

\IP Using wrapper scripts for build tools is considered a best practice by developers. We would like to show to what degree this manifests empirically by build success rates.

\RQ{RQ9}{What are the most common build failure reasons?}

In this question, we will divide the failed builds into a fixed set of categories using semi-automated analysis. The percentages of categories will be computed per JDK version and in total. Each category will be also qualitatively described using multiple specific examples of failures.

\IP If we find certain kinds of build failures are more prevalent, developers and tool designers can prioritize their efforts on fixing them. For instance, if we find dependency resolution problems are the most common, running a quick dependency scan could reveal many failures without running the full build process. \IR Researchers in the area of automated build repair will know on which problems to focus their approaches to fix the largest proportion of errors.

%%%%%%%%%%%%%%%%%%%%%%%%%%%%%%%%%%%%%%%%%%%%%%%%%%%%%%%%%%%%

\section{Dataset}
\label{s:dataset}

There exists a dataset containing local Java build results and logs \cite{Sulir20large}, but it was produced using only JDK 8 and four years later only with JDK 11. We are aware of datasets containing CI logs, such as TravisTorrent \cite{Beller17oops}, Continuous Defect Prediction \cite{Madeyski17continuous}, and LogChunks \cite{Brandt20logchunks}. However, in our study, we would like to focus on local builds, simulating an environment used by a developer or a researcher.

Since our goal is to acquire the full source code of the most up-to-date commit of many open-source projects, we will download them directly from the largest Git hosting provider, GitHub\footnote{\url{https://github.com}}. As we are not aware of any efficient and straightforward way to obtain a random sample of projects matching specific criteria using GitHub API, we will use the GitHub Search tool by Dabic et al. \cite{Dabic21sampling} for random sampling. The inclusion criteria are as follows:

\begin{description}[leftmargin=0cm]
\item[I1] The main programming language of the project, as determined by GitHub's algorithm, is Java.
\item[I2] A machine-recognizable open source License file is a part of the repository.
\item[I3] The repository has at least 10 stars. This is a limitation imposed by the used search tool. At the same time, it filters the data to include only engineered software projects and exclude toy/sample projects and homework with very high precision, although sacrificing recall \cite{Munaiah17curating}.
\item[I4] The project contains a Gradle, Maven, or Ant configuration file in its root directory. In Table~\ref{t:tools}, we list the recognized files. The tools are ordered in descending priority in case the project contains multiple configuration files. A rationale behind this is that newer, more modern build tools tend to replace the older ones whose settings can be left in the repository for compatibility.
\end{description}

\begin{table}
\caption{The Recognized Build Tools}
\label{t:tools}
\centering
\begin{tabular}{ll}
\toprule
Build System & File Name\\
\midrule
Gradle & build.gradle \\
~ & build.gradle.kts \\
Maven & pom.xml \\
Ant & build.xml \\
\bottomrule
\end{tabular}
\end{table}

We will also apply the following exclusion criteria:

\begin{description}[leftmargin=0cm]
\item[E1] Being a fork.
\item[E2] Having the same byte-for-byte files content despite not being marked as a fork, which happens in a small number of cases \cite{Lopes17dejavu}.
\item[E3] Utilizing the Android, Java Micro Edition (Java ME), or Java Native Interface (JNI) technologies. These technologies require the installation of separate toolchains, often under proprietary licenses. Therefore, we will focus on pure Java, excluding these separate ecosystems. To determine their presence, we will use the file name and content matching criteria from our previous paper \cite{Sulir20large}, described in Table~\ref{t:excluded}.
\end{description}

\begin{table}
\caption{Technologies Excluded from the Study}
\label{t:excluded}
\centering
\begin{tabular}{lll}
\toprule
Technology & File Name Pattern & File Content\\
\midrule
Android & *.java & import android. \\
~ & AndroidManifest.xml & --- \\
Java ME & *.java & import javax.microedition. \\
JNI & *.~\{c,cc,cpp,cxx\} & JNIEXPORT \\
\bottomrule
\end{tabular}
\end{table}

Criteria I4, E2, and E3 are not supported by the search tool or GitHub API. Therefore, we will obtain a complete list from the search tool, start downloading repositories in random order, and add the projects fulfilling all criteria into our sample until the limit of 2,500 is reached.

For each repository, we will create a shallow Git clone of its default branch, i.e., the copy only of the last commit. This will save resources and still prevent many failures of build tool plugins utilizing Git history. Git submodules (links to commits in other repositories) will not be downloaded since Git neither clones them by default nor warns a user about their presence.

As a form of a quality check, we will manually inspect a small number of projects and find whether they actually fulfill the inclusion and exclusion criteria. The complete source code of 2,500 projects along with the necessary metadata (including repository names, commit hashes, stargazer counts, detected build tools, etc.) will be archived and available as a part of the replication package.

%%%%%%%%%%%%%%%%%%%%%%%%%%%%%%%%%%%%%%%%%%%%%%%%%%%%%%%%%%%%

\section{Execution Plan}

In this section, we will describe the planned method of the execution of our study.

\subsection{Environment Preparation}

First, we will prepare eighteen Docker\footnote{\url{https://www.docker.com}} images containing the software necessary to run the build processes. Each image will contain:
\begin{itemize}
\item an up-to-date Linux distribution,
\item Git,
\item the latest minor release (update) of the given major JDK version,
\item the latest compatible version of Gradle and Maven,
\item Ant 1.9 and a compatible version of the Ivy dependency manager (its extension).
\end{itemize}

Apart from the JDK version, the images will be identical, with one exception. Since the latest versions of Gradle and Maven are incompatible with Java 6 and 7, we will include the last releases compatible with the given JDKs in these two containers. We do not consider this a major problem since Gradle projects often include a wrapper for downloading a specific Gradle version anyway, and Maven versions tend to differ only marginally in terms of features.

\subsection{Build Tool Execution}

\begin{table*}
\caption{The Commands Used to Execute The Build Tools}
\label{t:commands}
\centering
\begin{tabular}{llll}
\toprule
Tool & Wrapper & Installed & Command with Parameters \\
\midrule
Gradle & \texttt{./gradlew} & \texttt{gradle} & \verb/$gradle clean assemble --no-daemon --stacktrace --console=plain/ \\
Maven & \texttt{./mvnw} & \texttt{mvn} & \verb/$mvn clean package -DskipTests --batch-mode/ \\
Ant & \texttt{./antw} & \texttt{ant} & \verb/$ant clean; $ant jar || $ant war || $ant dist || $ant/ \\
\bottomrule
\end{tabular}
\end{table*}

The main part of the study will be executed fully automatically on a server. The building of every project/JDK combination will be performed in a separate, clean Docker container. We will extract the given project from the prepared archive. Based on the detected build tool, the corresponding command will be executed, as it is displayed in Table~\ref{t:commands}. If a wrapper script is present in the root directory, it will be run instead of the system-wide installed tool. The purpose of wrapper scripts is to download the correct build tool version, and it is a good practice to use them.

In the last column of Table~\ref{t:commands}, we can see the specific commands executed. First, possible leftovers after previous compilations are cleaned, in case they were committed to the repository by mistake. Then we attempt to compile the project and produce a distribution package, such as a JAR file. Since Ant does not have standardized target names, we try multiple common options before resorting to the execution of a default target.

Note that we exclude the running of tests if we have such an option. Although this might not represent typical users' behavior, it is usually possible to produce a binary package and inspect, run, or otherwise use it even if some of the tests fail. As a side effect, it makes the study less resource-intensive.

The rest of the command-line parameters represent options simplifying subsequent log analysis, such as the removal of ASCII color codes and the inclusion of stack traces.

To prevent infinite execution of stalled builds, each running time will be limited to one hour. We will capture the exit codes of the build processes, 0 meaning success, 124 timeout, and any other value failure. Complete standard output and error streams will be saved to files and become a part of the replication package.

\subsection{Data Analysis}
\label{s:analysis}

After the completion of all build process executions, we will analyze the produced data. To simplify the analysis, we will consider timed-out builds to be failed.

We will test the main hypothesis \textbf{H} statistically. First, we need to determine whether there is a trend at all, so we state the null and alternative hypothesis:
\begin{description}[leftmargin=0cm]
\item[H\textsubscript{0}] \textit{There is no monotonous trend in the proportion of build-failing projects in relation to the increasing JDK version numbers.}
\item[H\textsubscript{1}] \textit{There is a rising or falling trend in the proportion of build-failing projects in relation to the increasing JDK version numbers.}
\end{description}
We will test the hypothesis using the Mann-Kendall test with a significance level of 0.05. The Mann-Kendall test assesses whether a trend (increasing or decreasing) is present in the data. It is a nonparametric test, and the trend does not need to be linear.

If the computed p-value is less than 0.05, we will determine the direction of the trend. The Mann-Kendall test produces a value of tau, which can be used with an advantage. If tau is negative, the build-failing percentage is decreasing; if it is positive, the percentage is increasing.

The sole necessary input to answer \textbf{RQ1} (What proportion of projects is buildable using JDKs ranging from 6 to 23?) and selected questions about projects and JDKs (\textbf{RQ2}, \textbf{RQ3}, \textbf{RQ5}, \textbf{RQ6}) is a matrix of the build results. The columns of the matrix are JDK versions (6 to 23) and the rows are individual projects. Each element is a binary value: the given combination either passed (1) or failed (0).

The answer to question \textbf{RQ1} itself is a vector of percentages, computed by summing the columns, dividing them by the number of rows, and multiplying by 100. Questions \textbf{RQ2}, \textbf{RQ3}, and \textbf{RQ5} produce scalars that can be computed in a similarly trivial way.

\textbf{RQ4} is purely qualitative. The authors will manually inspect a sample of projects buildable under all JDKs. Build configuration files and portions of source code files will be inspected for signs of compatibility assurance. The findings will be summarized verbally.

To answer \textbf{RQ6} (Which JDK versions cause sudden drops in the passing rate?), we will produce a line plot. On the x-axis, there will be JDK versions from the oldest to the newest ones. The percentages of passing builds will be on the y-axis. An answer will be produced by a manual visual analysis of this plot. We consider automated pattern analysis unnecessary for such a small number of data points.

For \textbf{RQ7} (passing rates per JDKs and build tools), we need a mapping from a project to the detected build tool in addition to the build results matrix. The result will be a table with build tools (Gradle, Maven, Ant) as columns and individual JDK versions as rows, accompanied by a line graph. The columns in the table will be further split into two sub-columns ``wrapper''/``no wrapper'' to answer \textbf{RQ8}. Totals will be also computed for each column.

The purpose of \textbf{RQ9} is to divide the failed builds into categories. First, we will automatically extract essential information from the build logs. This process will be similar to our previous study \cite{Sulir16quantitative}. For both Maven and Gradle, we will extract the thrown exception name. If the exception is too general, we will append the goal name to it in the case of Maven; for Gradle, the task or plugin name will be appended. For Ant, we will extract only the last executed target name. As a result of this phase, each build will have an error type assigned, e.g., ``MojoFailure:maven-compiler-plugin''.

Then two authors will manually map such error types to one of a fixed set of categories, i.e., perform closed qualitative coding. Their assignments will be compared and discrepancies will be resolved by discussion. The list of possible categories was inspired by selected items from two build failure taxonomies by Vassallo et al. \cite{Vassallo17tale} and Lou et al. \cite{Lou20understanding}: build file parsing, dependency resolution, resource processing, compilation, documentation, extra plugin execution, packaging, timeout, and other/unknown.

For the ``timeout'' category, no log analysis will be necessary, as all timed-out builds will be assigned to it automatically. Due to a large expected number of error types, the special category ``other/unknown'' will be automatically assigned to error types occurring too infrequently. This category will be also used for error types where no agreement would be reached by discussion.

In justified cases, it will be possible to make small modifications to the list of categories after the agreement of all authors.

After finishing the assignment, the most frequently occurring failure categories will be reported for individual JDKs and in total.

\subsection{Quality Checks} \label{s:quality}

As one form of quality checks, unit tests will be written for some parts of the automated scripts where it is suitable. Furthermore, the source code of the scripts written by one author will be reviewed by another author to check for possible flaws and bugs as necessary.

On a larger scale (comparable to systems tests), we will implement a kind of positive control for the main hypothesis \textbf{H} and all research questions where it is relevant (\textbf{RQ1}--\textbf{RQ3}, \textbf{RQ5}, \textbf{RQ7}, and \textbf{RQ8}). A small number of build scripts that either always fail, always pass, or pass only for a specific Java version will be created. Then we will manually calculate or determine the results on this control set and compare them with the results computed by automated scripts.

%%%%%%%%%%%%%%%%%%%%%%%%%%%%%%%%%%%%%%%%%%%%%%%%%%%%%%%%%%%%

\section{Threats to Validity}

In this section, we will mention selected threats to validity that are apparent before executing the study. They are categorized according to Wohlin et al. \cite{Wohlin12experimentation}.

\subsection{Internal Validity} \label{s:internal}

For Gradle and Maven, we use standard and universally respected command-line options to skip the execution of tests. However, Ant scripts are more free-form. We will try to exclude tests by specifying target names that usually do not execute them (jar, war, dist), but there is no guarantee. More importantly, the default target, used as a fallback, executes all top-level tasks unless defined otherwise.\footnote{\url{https://ant.apache.org/manual/using.html}} We consider this a limitation of a rather archaic build tool that the end user whose behavior we try to imitate will encounter too.

Another problem may be caused by the project language detection. We utilize the information provided by the GitHub API, which uses file extensions to determine the language and counts the sizes of files to find the dominant language of the project. This means that a project may include also languages other than Java. In our study, however, we are only interested in the buildability of the Java part of the program, so the presence of other components should not affect it in most cases, especially after excluding JNI projects.

Current versions of Maven and Gradle do not support the two oldest studied Java releases (Java 6 and 7). Potential solutions are:

\begin{enumerate}
    \item Evaluate different combinations of Java and tools versions.
    \item Use the same version of the tools everywhere, but run them using an additional compatible version of Java while performing the build itself using the specified JDK configured by the toolchain mechanism.
    \item For Java 6 and 7, use the last versions of Maven and Gradle compatible with them.
\end{enumerate}

The first option was rejected because it would introduce other variables, making the execution of the study practically unfeasible and the interpretation of the results extremely complicated. Due to the variety of build configurations, the toolchain mechanism might not be respected during builds, which would cause hard-to-find discrepancies in the results. Therefore, we selected the third option. Gradle projects tend to include a wrapper script that makes the installed version irrelevant, and Maven stayed at the same major version, 3, for the last decade, with very few breaking changes. This behavior also most faithfully represents possible users' environments, thus improving the external validity.

\subsection{External Validity}

The research is limited by the project selection criteria. We decided to exclude projects requiring other toolchains in addition to the JDK --- Android, Java ME, and JNI. Their inclusion could make the results more general and represent a greater variety of projects. On the other hand, it would introduce additional dependencies and their versions may influence the results, threatening the internal validity of the study. Furthermore, some of them have proprietary licenses prohibiting unlimited redistribution, which would cause legal problems with the replication package.

Another limitation is requiring project repositories to have at least 10 stars. While this criterion allows us to exclude toy/sample projects with high precision, it has a low recall~\cite{Munaiah17curating}, thus excluding also a lot of engineered software projects. This criterion is a consequence of the used search tool. We could consider downloading all necessary metadata from GitHub and sample projects locally. This approach, however, might not be feasible because of the limitations exposed by the GitHub API.

\subsection{Reliability} \label{s:reliability}

The whole study except \textbf{RQ4}, the visual analysis part of \textbf{RQ6}, and the category tagging part of \textbf{RQ9} will be performed fully automatically by scripts. The complete dataset and all scripts will be archived in a publicly available replication package. As for \textbf{RQ9}, we will not calculate inter-rater reliability, but resolve all discrepancies by discussion instead. All cases when the discussion will not lead to a mutual agreement will be marked as ``other/unknown''. This will ensure a high degree of reliability at the expense of having a small part of the dataset unlabeled. We will expect such situations to occur mainly in cases when an error type does not contain sufficient information to distinguish between multiple diverse behaviors that lead to this error.

\begin{acks}
This work was supported by project VEGA No. 1/0630/22 Lowering Programmers' Cognitive Load Using Context-Dependent Dialogs.
\end{acks}

\balance

\bibliographystyle{ACM-Reference-Format}
\bibliography{esem}

\end{document}